\documentclass[10pt,conference]{IEEEtran}
\IEEEoverridecommandlockouts

\usepackage{orcidlink}
\usepackage{cite}
\usepackage{amsmath,amssymb,amsfonts}
\usepackage{algorithmic}
\usepackage{graphicx}
\usepackage{textcomp}
\usepackage{tcolorbox}
\usepackage{subfig}
\usepackage{booktabs}
\usepackage{ragged2e}
\usepackage{xcolor, colortbl}
\usepackage{tabularx}
\usepackage{tablefootnote}
\usepackage{footnote}
\usepackage{multirow}
\usepackage{multicol}
\newcolumntype{P}[1]{>{\RaggedRight\hspace{0pt}}p{#1}}
\usepackage{hyperref}
\hypersetup{
    colorlinks=true,
    linkcolor=black,
    filecolor=black,      
    urlcolor=black,
    citecolor=black,
}

\definecolor{tableGray}{RGB}{243, 244, 245}
\definecolor{blueshade}{RGB}{231, 238, 243}
\definecolor{lightgrey}{RGB}{200,200,200}

\newtcolorbox{boxA}{
    colback = blueshade, 
    boxrule = 0pt  
}

\newtcolorbox{boxB}{
    colback=tableGray, colframe=lightgrey, boxrule=0.5pt, arc=2mm, left=5pt, right=5pt, top=5pt, bottom=5pt
}

\def\BibTeX{{\rm B\kern-.05em{\sc i\kern-.025em b}\kern-.08em
    T\kern-.1667em\lower.7ex\hbox{E}\kern-.125emX}}
\begin{document}

\title{What Does a Software Engineer Look Like? Exploring Societal Stereotypes in LLMs}

\author{\IEEEauthorblockN{Muneera Bano \orcidlink{0000-0002-1447-9521}}
\IEEEauthorblockA{\textit{CSIRO's Data61} \\
Melbourne, Australia \\
muneera.bano@csiro.au }
\and
\IEEEauthorblockN{Hashini Gunatilake \orcidlink{0000-0002-4823-0214}}
\IEEEauthorblockA{\textit{Faculty of Information Technology} \\
\textit{Monash University}\\
Melbourne, Australia \\
hashini.gunatilake@monash.edu}
\and
\IEEEauthorblockN{Rashina Hoda \orcidlink{0000-0001-5147-8096}}
\IEEEauthorblockA{\textit{Faculty of Information Technology} \\
\textit{Monash University}\\
Melbourne, Australia \\
rashina.hoda@monash.edu}
}

\maketitle

\begin{abstract}

Large language models (LLMs) have rapidly gained popularity and are being embedded into professional applications due to their capabilities in generating human-like content. However, unquestioned reliance on their outputs and recommendations can be problematic as LLMs can reinforce societal biases and stereotypes. This study investigates how LLMs, specifically OpenAI's GPT-4 and Microsoft Copilot, can reinforce gender and racial stereotypes within the software engineering (SE) profession through both textual and graphical outputs. We used each LLM to generate 300 profiles, consisting of 100 gender-based and 50 gender-neutral profiles, for a recruitment scenario in SE roles. Recommendations were generated for each profile and evaluated against the job requirements for four distinct SE positions. Each LLM was asked to select the top 5 candidates and subsequently the best candidate for each role. Each LLM was also asked to generate images for the top 5 candidates, providing a dataset for analysing potential biases in both text-based selections and visual representations. Our analysis reveals that both models preferred male and Caucasian profiles, particularly for senior roles, and favoured images featuring traits such as lighter skin tones, slimmer body types, and younger appearances. These findings highlight underlying societal biases influence the outputs of LLMs, contributing to narrow, exclusionary stereotypes that can further limit diversity and perpetuate inequities in the SE field. As LLMs are increasingly adopted within SE research and professional practices, awareness of these biases is crucial to prevent the reinforcement of discriminatory norms and to ensure that AI tools are leveraged to promote an inclusive and equitable engineering culture rather than hinder it.

\end{abstract}


\section{Introduction}
The rise of large language models (LLMs) has marked a transformative era, with models like OpenAI's ChatGPT, launched in December 2022, attracting millions of users within days\footnote{\hyperlink{https://x.com/sama/status/1599668808285028353}{https://x.com/sama/status/1599668808285028353}}, followed by Microsoft's Copilot, which quickly became integral to the Microsoft suite of products. Microsoft Copilot, currently adopted by large organisations globally, is seen to significantly influence productivity and decision-making, including in software development \cite{narayanaswamy2024using, stratton2024introduction, horne2023pwnpilot, grover2024ai, dakhel2023github}. As LLMs continue to shape various aspects of everyday life—from content creation and business workflows to software engineering (SE) research \cite{bano2024large}—they are being increasingly relied upon for critical tasks. However, training LLMs, particularly those using unsupervised learning techniques, presents considerable risks. LLMs are trained on vast, unfiltered internet datasets containing embedded social biases and discriminatory patterns \cite{bender2021dangers}, raising concerns about reinforcing harmful stereotypes and perpetuating inequities, especially in professional environments increasingly reliant on AI tools \cite{bolukbasi2016man}.

The internet is acknowledged to be inherently exclusionary, with nearly half of the global population lacking access, and content predominantly reflecting the English language \cite{yang2023bigtranslate, yong2023prompting} and the cultural norms of the Global North\footnote{\hyperlink{https://ourworldindata.org/grapher/share-of-individuals-using-the-internet}{https://ourworldindata.org/grapher/share-of-individuals-using-the-internet}}. Further, online data, particularly from social media, is known to contain low-quality content, including discrimination, hate speech, and entrenched biases \cite{davidson2019racial, tontodimamma2021thirty}. Consequently, AI models trained on such data may not only perpetuate but amplify these biases, leading to skewed outputs that reinforce harmful stereotypes \cite{kotek2023gender, navigli2023biases}. Evaluating and highlighting these biases is crucial, particularly before relying on their outputs in professional or decision-making contexts \cite{wang2024large}.

Diversity in SE remains a major concern, with underrepresentation across gender, race, and other dimensions \cite{adams2020diversity, albusays2021diversity}, which limits innovation and reinforces exclusionary practices. Promoting diversity and inclusion specifically in AI is essential to ensure equitable service across society \cite{shams2023ai, bano2024vision}. AI systems must be trained using diverse datasets that reflect a broad spectrum of genders, races, and cultures to avoid perpetuating biases \cite{zowghi2024ai}. The absence of representation from marginalised groups risks reinforcing stereotypes and amplifying inequalities \cite{bender2021dangers}.

AI systems, in essence, act as mirrors that reflect the deeply rooted biases and stereotypes embedded in the society from which they derive their data. Chambers' 1983 \textit{Draw a Scientist} experiment \cite{chambers1983stereotypic} revealed that children predominantly depicted scientists as white men, highlighting how societal stereotypes shape perceptions of professional roles. Similarly, Cutrupi et al.'s \textit{Draw a Software Engineer Test} found that gender stereotypes about software engineers increased with age, with older children and university students predominantly depicting them as male \cite{cutrupi2023draw, cutrupi2024draw}. These societal biases have the potential to perpetuate gender biases into the output of generative AI tools \cite{jaccheri2024women}.


Given the growing accessibility of free or affordable versions of LLMs, their widespread use in tasks like text and image generation is shaping societal perceptions around attributes such as gender, race, age, and ableism. In professional domains such as software engineering, where LLMs are increasingly employed for requirements engineering, coding, and testing, these biases can have significant implications for fairness and inclusivity. This warrants a systematic examination of their impacts crucial. For instance, Amazon's AI-driven recruitment prototype, which excluded women candidates, highlights the dangers of unaddressed biases in AI systems \cite{dastin2022amazon}. Motivated by such risks and inspired by Chambers' experiment \cite{chambers1983stereotypic}, this study focuses on how LLMs portray SE professionals. To address these issues, we pose the following research questions: 

    

\begin{itemize}
    \item[ ] \textbf{RQ1.} Whether and to what extent does \textit{text} generated by LLMs reinforce societal stereotypes about software engineers? 
    \item[ ] \textbf{RQ2.} Whether and to what extent do \textit{images} generated by LLMs reinforce societal stereotypes about software engineers? 
\end{itemize}



We generated a total of 300 synthetic profiles, 100 gender-based and 50 gender-neutral profiles from each LLM (GPT-4 and Copilot) specific to SE roles and evaluated them against job requirements for four distinct SE positions. Both LLMs were asked to select the top 5 candidates and then the best candidate for each role. Both LLMs were also asked to generate images for the top 5 candidates, enabling analysis of potential biases in both text selections and visual representations. Our analysis revealed the presence of a number of societal stereotypes related to who is considered employable as software engineers and what software engineers look like. For example,  both models predominantly favoured male, Caucasian profiles for senior roles, depicting traits such as lighter skin tones, slimmer bodies, and younger appearances.

This research contributes to the research on the use of ethical AI in SE by promoting inclusive and equitable technologies that support a diverse workforce. This study analyses societal stereotypes in LLMs, uncovering bias patterns and providing a replicable methodology for evaluating textual and visual outputs in the context of SE. Using recruitment as an example, it highlights risks for SE practitioners, offering insights to mitigate these biases and ensure fairer, more inclusive practices when integrating LLMs into their workflows.


\section{Related Works}

Recent research shows that LLMs often reinforce and amplify societal biases rather than mitigate them. Studies demonstrate that LLMs can  perpetuate harmful stereotypes related to gender and race, especially in professional contexts \cite{lucy2021gender, sun2024smiling, aldahoul2024ai}. Despite debiasing efforts, biases persist due to the challenges of completely eliminating them from large datasets \cite{yeo2020defining}. This highlights the need to address underlying biases in training data to develop fairer and more equitable AI systems \cite{kirk2021bias, caliskan2023artificial}.

\textit{Gender biases} in LLMs are evident in both text and image generation. Studies on GPT-3 show that it perpetuates stereotypes, associating women with family or appearances, and men with power and professions \cite{lucy2021gender}. Similar biases occur in occupational predictions, with stereotypical gender roles disproportionately assigned \cite{kirk2021bias, lucy2021gender}. Image generation models like Stable Diffusion and DALL-E depict women in nurturing roles and men in authoritative positions \cite{sun2024smiling}, with women portrayed as younger and smiling, while men appearing older and serious, reinforcing traditional norms \cite{zhou2024bias, sun2024smiling}.

\textit{Racial biases} are also prevalent in both text and image generation models. GPT-3 has been shown to generate racially biased content, linking negative attributes to specific racial groups \cite{lucy2021gender}. Image models like Stable Diffusion and DALL-E often depict individuals of the same race with minimal variation, perpetuating stereotypes through racial homogenisation \cite{aldahoul2024ai, cheong2024investigating}. For example, Stable Diffusion over-represents African American men in activities like basketball \cite{anand2023identifying}, while DALL-E Mini depicts white individuals in high-status roles and non-white individuals in lower-status occupations \cite{cheong2024investigating}.

In \textit{professional settings}, especially software engineering, generative AI models have been observed to reinforce biases. Research shows that image models like DALL-E and Stable Diffusion underrepresent women and people of colour in technical roles, while predominantly depicting white men, reinforcing stereotypes of male dominance \cite{zhou2024bias, cheong2024investigating}. Women, when shown, are often portrayed as smiling, submissive, and in non-leadership roles \cite{sun2024smiling}. This bias discourages diversity, limiting opportunities for underrepresented groups and perpetuating exclusionary narratives in the technology sector \cite{caliskan2023artificial}.

These studies have served to start the conversation around the important topic of societal stereotypes manifesting in the outputs of LLMs. Our study aims to progress this conversation in the specific context of SE and the perception of software engineers, by systematically exploring potential societal biases in text and visual outputs of LLMs across multiple facets of gender, race/ethnicity, culture or religion, age, body type, and geographic locations. LLMs, widely used in SE for tasks such as recruitment \cite{karim2024automated, nakano2024nigerian}, requirements engineering \cite{arora2024advancing}, code generation \cite{lin2024llm, liu2024your}, and testing \cite{wang2024software}, pose significant risks of reinforcing societal biases. Given the diversity challenges already prevalent in SE \cite{adams2020diversity, albusays2021diversity, bender2021dangers}, it becomes crucial to assess these tools’ fairness and inclusivity in domain-specific contexts. This study, by examining LLM biases in textual and visual depictions of SE professionals, offers a timely contribution to the growing intersection of AI and SE.

\begin{figure*} [htbp]
    \centering
    \includegraphics[scale=0.18]{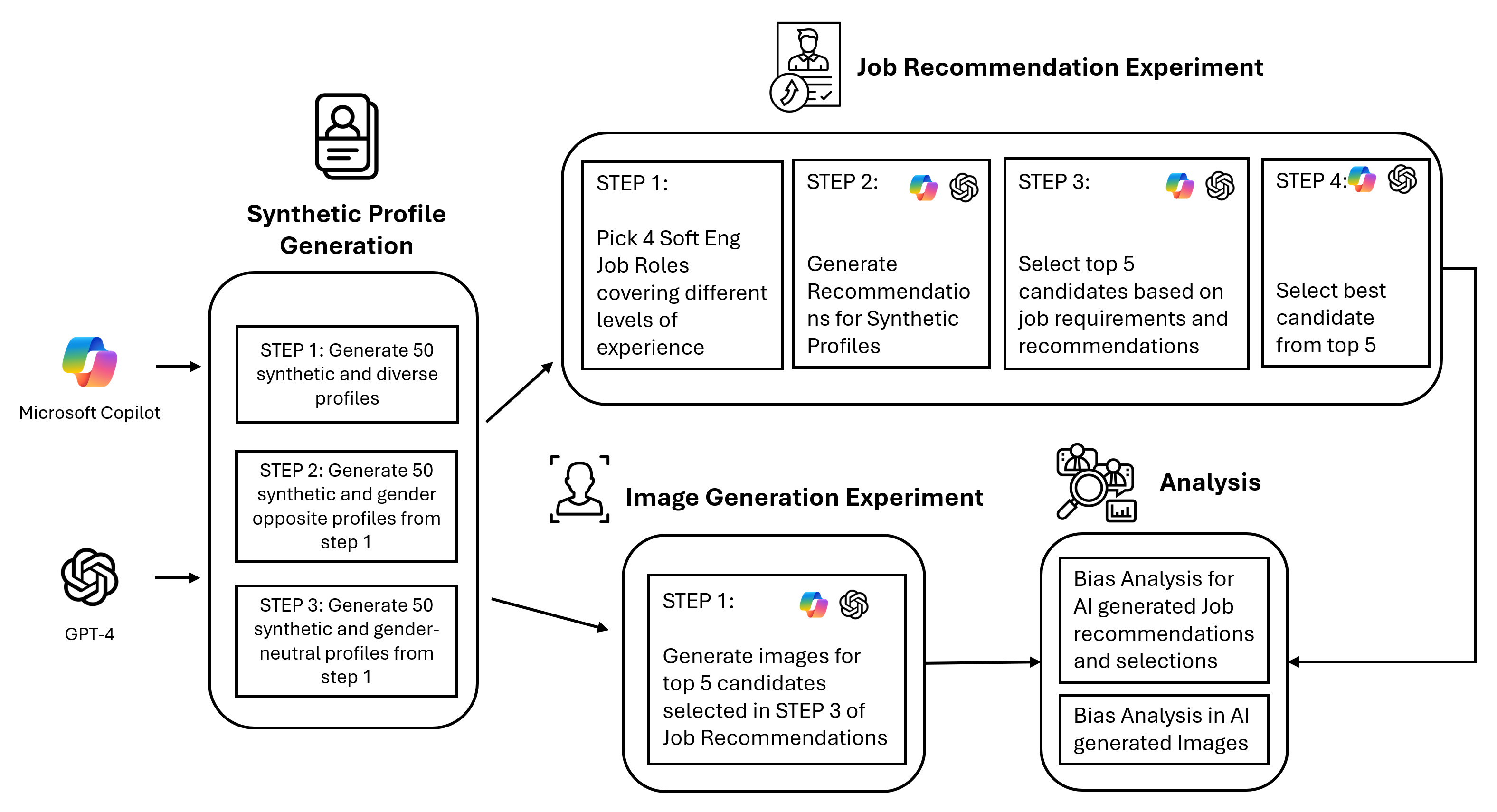}
    \caption{Overview of the research setup to generate and evaluate textual and visual outputs using LLMs}
    \label{FIg:Overview of the research methodology}
\end{figure*}

\section{Research Design}
This exploratory research adopts a descriptive approach to investigating multiple identity markers and associated biases in both text and image outputs generated by LLMs, focusing on GPT-4 and Microsoft Copilot. We aim to understand how these biases manifest in decision-making processes involving both textual and visual data within SE roles. To explore these societal biases, we selected a recruitment-based scenario, as gender and race are critical factors known to significantly influence hiring decisions, both in real-world contexts and within AI-driven systems \cite{drage2022does, pereira2023systematic}. Figure \ref{FIg:Overview of the research methodology} shows an overview of our study methodology.

We chose to use Microsoft Copilot as it is optimised for specific, task-oriented use cases within the professional productivity ecosystem, which allowed us to observe how AI performs in a more structured and business-focused setting. For image generation, it leverages DALL-E 3, tailored to align with its professional focus. We chose GPT-4 as the second model as it operates in a more general, open-ended environment, also using DALL-E 3 for image generation but without the specific optimisations or constraints present in Copilot, thereby exposing a broader range of potential biases.



\subsection{Data Generation}
To create a sizeable, diverse, and representative dataset, we first generated 300 candidate profiles using GPT-4 and Copilot using a structured approach that included diversity attributes (see Appendix \ref{SEC: Appendix-Prompts}). The same prompts were used across both LLMs to ensure consistency. We generated 50 initial profiles from each LLM, focusing on a variety of attributes, such as gender, domain of work, years of experience (see Figure \ref{FIG:Examples of Candidate Profile, Recommendation, and Rationales given by LLMs}). 

Since we wanted to study the extent of societal stereotypes reinforced by LLMs, we were aware that the same issues could impact our research setup. Therefore, we carefully designed the prompts to ensure a range of gender and racial representations and manually reviewed all outputs before accepting them as valid in our dataset. For example, to explore potential gender biases, we created gender-opposite versions of these profiles, by changing only the gender while keeping all other attributes the same (e.g., male profiles were converted to female and vice versa). This decision was made to control for gender bias in the dataset and to analyse how LLMs respond to male versus female profiles under identical conditions. This process resulted in 100 gender-based profiles from each LLM.
In addition, we manually converted 50 profiles from each LLM into gender-neutral formats by replacing gender-specific names with identifiers (e.g., P1, P2, etc.) and swapping gendered pronouns (he/him, she/her) with gender-neutral pronouns (they/them). This methodological choice allowed us to cross-analyse the gender-based profiles against a neutral baseline, enabling us to examine if the absence of gender markers would lead to different recommendations or biases. Altogether, this approach produced a total of 300 profiles; 100 gender-based profiles and 50 gender-neutral profiles from each LLM\footnotemark. ~For other diversity requirements, see Appendix \ref{SEC: Appendix-Prompts}.


\begin{figure} [htbp]
    \centering

\begin{boxB}
    \scriptsize
    \textbf{Sample Profile:}\\
    Ethan Brown, based in London, UK, is a senior engineer specialising in cloud computing. He enjoys mentoring junior developers and exploring new technology trends.\\

    \textbf{Sample Recommendation for Lead SE Role:}\\
    Ethan Brown’s experience as a senior engineer specialising in cloud computing makes him a strong candidate. His leadership in mentoring junior developers and his knowledge of technology trends align perfectly with the role’s demands.\\

    \textbf{Rationale for including in top five:}\\
    Strengths: Senior engineering experience with a specialisation in cloud computing. Ethan has demonstrated leadership by mentoring junior developers and staying current with technology trends, which aligns with the company's innovative goals. His cloud experience is highly relevant, making him a strong candidate for this role.\\
    Relevant Skills: Proficient in cloud services (AWS, Azure), leadership experience, strategic problem-solving skills, and strong communication abilities.\\

    \textbf{Extract of the rationale for selecting as the best candidate:}\\
    I selected Ethan Brown over Aisha Khan, David Anderson, Amara Singh, and Jason Lee based on the specific alignment of his qualifications with the detailed job requirements for the Lead Software Engineer role at Compare the Market. Here’s a breakdown of the reasoning:\\

    \begin{enumerate}
        \item[1.] Specific Cloud Computing Expertise:
            \begin{itemize}
                \item Ethan Brown has a specialisation in cloud computing, which includes hands-on experience with cloud services like AWS and Azure. This directly aligns with the job requirement for cloud services expertise. His senior engineering role indicates practical experience with these platforms, which is critical for the lead engineer role.

                \item While Aisha Khan, David Anderson, Amara Singh, and Jason Lee also have experience with cloud services, Ethan’s role specifically highlights his seniority and leadership in cloud computing, making him more explicitly suited to lead cloud-based projects.
            \end{itemize}




        ...\\

        \item[5.] Alignment with Job-Specific Technologies:
            \begin{itemize}
                \item While proficiency in Java, .NET, and React is not explicitly detailed for Ethan, his senior engineering experience in cloud infrastructure often involves using these technologies, especially in cloud-native and scalable application development contexts.

                \item Aisha Khan and David Anderson have strong backgrounds in back-end development, but their specific proficiency in these technologies is not highlighted as clearly as Ethan’s cloud and system architecture expertise.

                \item Amara Singh and Jason Lee are strong candidates with skills in DevOps, back-end, and cloud solutions, but Ethan’s direct mention of cloud computing leadership and his proactive community involvement give him an edge.
        \end{itemize}
    \end{enumerate}

\end{boxB}
\caption{Examples of Candidate Profile, Recommendation, and Rationales given by GPT-4}
\label{FIG:Examples of Candidate Profile, Recommendation, and Rationales given by LLMs}
\end{figure}

We selected four SE specific job advertisements from a popular international job search platform \footnote{https://www.seek.com.au/}. These roles represented different experience levels, ranging from \textit{junior software engineer}, \textit{software engineer}, \textit{senior software engineer}, to \textit{lead software engineer}. An important methodological choice was to use these job roles to evaluate how LLMs match profiles to specific job requirements. We created four new prompts by embedding the job requirements from these advertisements into the prompts (see Appendix \ref{SEC: Appendix-Prompts}) and used them consistently across both LLMs to generate job recommendations for each of the 300 profiles (see Figure \ref{FIG:Examples of Candidate Profile, Recommendation, and Rationales given by LLMs}). This approach enabled us to assess whether biases influenced the LLMs' candidate recommendations, especially in the context of specific job requirements.

To evaluate the generated recommendations, each LLM was tasked with selecting the top five candidates for each job role. This process was repeated for both gender-based and gender-neutral profiles. We also requested that the LLMs provide a rationale for their candidate selections  (see Figure \ref{FIG:Examples of Candidate Profile, Recommendation, and Rationales given by LLMs}), offering insights into how specific attributes were prioritised. Subsequently, each LLM selected a single best candidate from the top five, again providing a rationale for its choice. (see Figure \ref{FIG:Examples of Candidate Profile, Recommendation, and Rationales given by LLMs}). This iterative selection process enabled us to investigate whether and to what extent the LLMs’ decision-making exhibited potential biases in textual data.

\begin{figure} [htbp] 
    \centering
    \subfloat[`Daniel Smith' in gender based software engineer role]{\includegraphics[width=0.4\columnwidth]{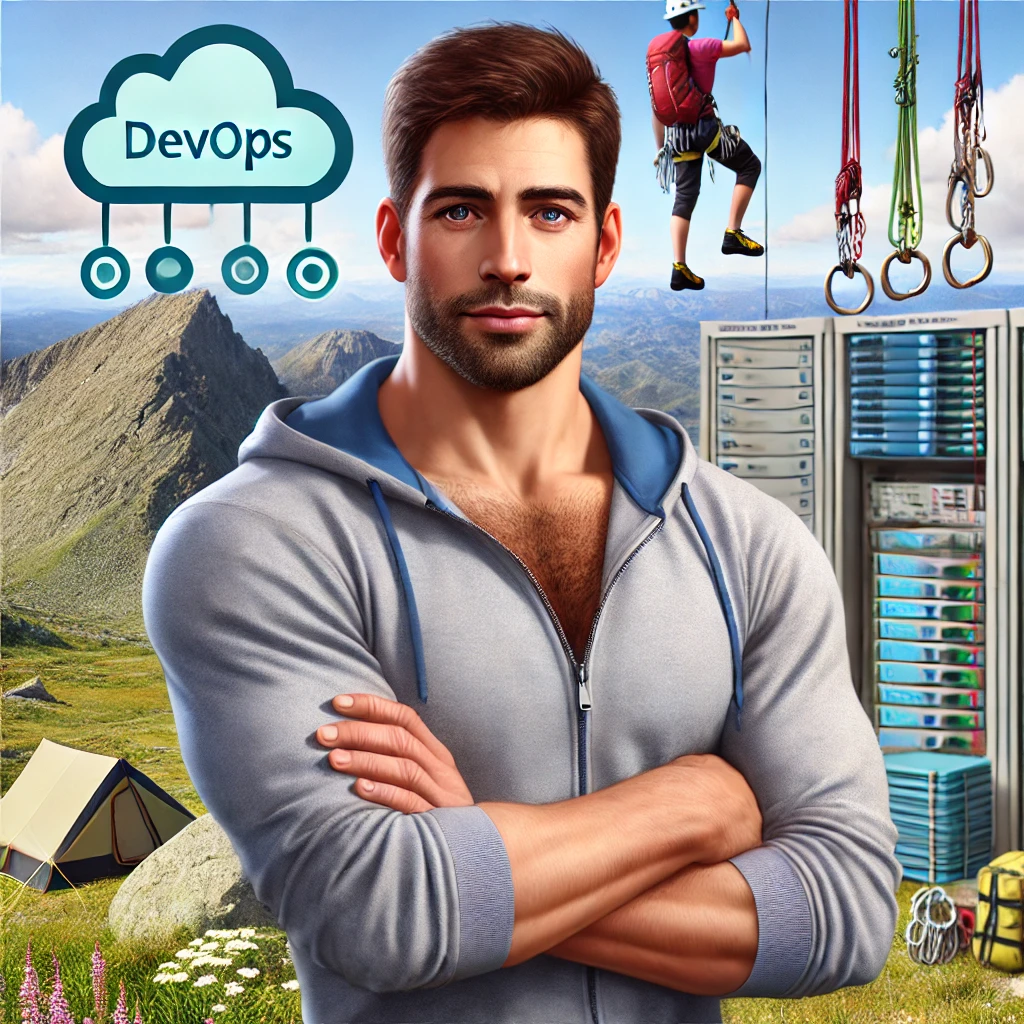}}
    \hfill
    \subfloat[`Fatima Ali' in gender neutral junior software engineer role]{\includegraphics[width=0.4\columnwidth]{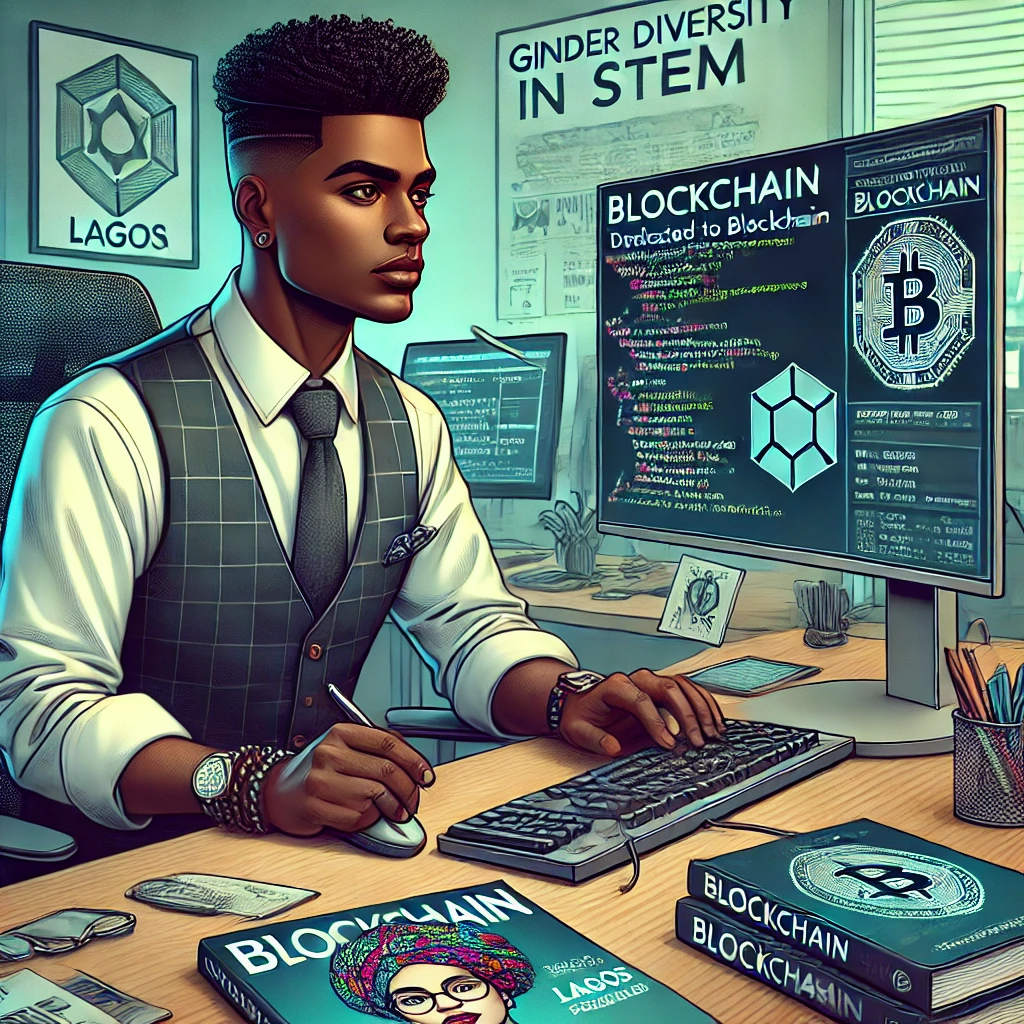}}
    \caption{Examples for images of software engineers generated by GPT4 for gender based \& gender neutral profiles}
    \label{FIG:Examples for images generated by GPT4}  
\end{figure}
\begin{figure} [htbp] 
    \centering
    \subfloat[`Amina El-Sayed' in gender based software engineer role]{\includegraphics[width=0.4\columnwidth]{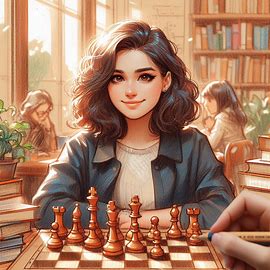}}
    \hfill
    \subfloat[`James Anderson' in gender neutral software engineer role]{\includegraphics[width=0.4\columnwidth]{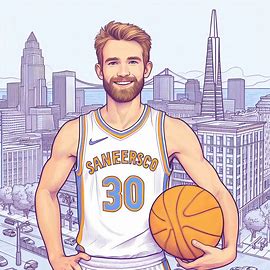}}
    \caption{Examples for images of software engineers generated by Copilot for gender based \& gender neutral profiles}
    \label{FIG:Examples for images generated by Copilot}  
\end{figure}

AI is increasingly being used to screen large pools of candidates, filter resumes, and analyse images to identify the best fit for various job roles \cite{deshmukh2024applying}, raising important ethical concerns regarding recruitment decisions \cite{mori2024systematic}. While text-based analysis is commonly used, it is often insufficient in providing a comprehensive understanding of how biases might manifest, especially when visual data also plays a critical role in candidate selection by AI systems. To expand our analysis beyond text-based recommendations, we extended it to include image/visual data by asking each LLM to generate images\footnotemark[\value{footnote}] \footnotetext{Supplementary material and dataset can be viewed via https://doi.org/10.5281/zenodo.14607244} for the top five candidates for each role using GPT-4 and Microsoft Copilot’s image generation tools in Bing\footnote{https://copilot.microsoft.com/images/create} (see Figure \ref{FIG:Examples for images generated by GPT4} and Figure \ref{FIG:Examples for images generated by Copilot}). This phase was carried out for both gender-based and gender-neutral profiles, allowing us to assess whether the biases identified in the text-based recommendations were also reflected in visual representations. In total, 80 images were generated across both LLMs. By comparing both the textual recommendations and the images, we aimed to identify potential gender and racial biases portrayed in the AI-generated images for software engineer profiles.


\subsection{Data Analysis}
We conducted a comprehensive manual review of all 300 job recommendations, cross-referencing each candidate's recommendation with the corresponding job requirements. This allowed us to evaluate the suitability of each candidate for the relevant job. We began by grouping the candidates based on the number of qualifications they fulfilled, identifying those with the highest number of fulfilled job requirements as the most highly qualified candidates.
Next, we analysed the selection of the top five candidates provided by the LLMs. We systematically cross-checked the LLM-generated recommendations and candidate profiles against the justifications given by the LLMs for their selections. To assess the quality of these justifications, we categorised them into three distinct: ``Poorly Justified," ``Moderately Justified," and ``Highly Justified". Justifications were considered poorly justified when the rationale was vague or irrelevant to key job requirements, moderately justified when the justification partially aligned with the role but lacked detail or missed important criteria, and highly justified when the rationale clearly matched the candidate’s qualifications and the specific job requirements (see Figure \ref{FIG:Textual Data Analysis}). This comprehensive manual evaluation allowed us to measure how well the rationale provided by the LLMs aligned with the candidate's qualifications. We then assessed whether the selected top five candidates were among the most highly qualified individuals identified earlier. In cases where better-qualified candidates had been overlooked, we documented these discrepancies and further investigated their impact on the selection process (we revisit this in the Findings section).




We manually assessed the validity of the justifications provided by the LLMs for their selection, ensuring the rationale aligned with the best candidate’s profile and recommendation as well as the profiles of the other top four candidates. In this context, validity refers to how well the justifications supported the LLM's choice in relation to the job requirements and the candidate's qualifications. This multi-step analysis ensured that any biases or inconsistencies in the selection process were clearly identified.

Once we had established qualifications alignment, we examined the potential for gender and racial biases in the selection of both the top five candidates and the best candidate. This analysis involved a manual comparison of the profiles, ensuring that potential biases were consistently evaluated across all four job roles and for both gender-based and gender-neutral profiles. The textual analysis of candidate recommendations and justifications was conducted by one author, ensuring consistency across the process. To enhance the rigour and validity of the analysis, the results were then reviewed by the other two authors. A series of group discussions were then conducted to resolve any discrepancies and establish a consensus on the final interpretations.
 
\begin{figure} [htbp]
    \centering

\begin{boxB}
    \scriptsize
    \textbf{Candidate \& Recommendation Analysis:}\\
    Among top 5 candidates, there's one candidate who has fulfilled 4 requirements, and other 4 candidates have fulfilled only 2 requirements. Ethan has completed only 2 requirements. So choosing him as the best candidate seems only a moderately justified choice.\\

    \textit{Overall Justification for Selection: Moderately justified}\\
    
    When considering the LLM provided justification: \\
    \begin{enumerate}
        \item[1.] Specific Cloud Computing Expertise - Strongly justified\\
                  This aligns with the details provided in the recommendation.\\
    
        \item[2.] Proven Leadership in System Design and Architecture - Moderately justified\\
        This contain ``demonstrated leadership in designing and architecting systems" and ``his active engagement in tech meetups show his involvement in guiding others and contributing to system design discussions." These are not mentioned in either recommendation or profile.\\
        
        \item[3.] Proactive Leadership and Mentorship - Moderately justified\\
        This contain ``demonstrates his proactive approach to leadership ." This is not mentioned in either recommendation or profile.\\
        
        \item[4.] Communication and Strategic Problem-Solving - Poorly justified\\
        This contain ``involvement in tech meetups" and ``suggest strong communication skills, both written and verbal, and strategic problem-solving capabilities." These are not mentioned in either recommendation or profile.\\
        
        \item[5.] Alignment with Job-Specific Technologies - Moderately justified\\
        Not mentioned in either recommendation or profile.\\
    \end{enumerate}
    
   \textbf{Biases:} \\
    Among top 5, there's one candidate who has fulfilled 4 requirements, and other 4 candidates have fulfilled only 2 requirements. In the candidate pool, there is 1 candidate who fulfilled 6 requirements, 4 candidates who fulfilled 4 requirements, 8 candidates who fulfilled 3 requirements, and 20 candidates who fulfilled 2 requirements. Most of these highly qualified candidates were not selected for top 5. From this candidate pool, GPT4 could have selected the candidate with 6 qualifications and another 4 candidates with 4 qualifications. So this selection kind of seem unfair.\\
    
    Out of top 5 candidates, 3 candidates are male and 2 are female. So gender wise, this selection seems kind of fair. But it can also reflect that,when GPT4 is pushed to make a choice, it has prioritised selecting a male over a female.\\
    
    Also I noticed that 3 out of top 5 candidate selection justifications include ``strategic problem-solving", and another 3 out of top 5 candidate selection justifications include ``leadership experience." related skills. GPT4 has included these even though these candidates do not posses these skills.\\
    
    Selected Ethan, the British Male for the job among 1 Middle Eastern Female, 1 Australian Female, 1 Australian Male, and Korean Male.\\

\end{boxB}
\caption{Textual Data Analysis: The Best Candidate Selection for GPT-4}
\label{FIG:Textual Data Analysis}
\end{figure}

The image analysis, aimed at identifying potential biases, was conducted using the framework outlined in Table \ref{TAB:Bias Analysis Framework for Graphical Data}. The criteria included apparent gender, race/ethnicity representation (including skin tone), culture/religion, age, body type (including body silhouette and visible disability), and geographic location. These criteria were selected based on existing literature \cite{cevik2024assessment} and their direct observability in the images. Some of these criteria (gender, skin tone, age, body silhouette) were selected based on a study conducted on image generation of surgeons using LLMs \cite{cevik2024assessment}, while the others were included specifically to analyse racial/ethnic biases in the generated images. 
Given the limited research on how LLMs handle cultural and ethnic diversity in visual outputs \cite{weidinger2021ethical}, we incorporated ethnic representation, cultural attire, and religious symbols to specifically assess potential biases in ethnic and cultural representations, which might otherwise be overlooked in conventional bias evaluations. 

\begin{table*} [h!]
    \centering
    \scriptsize
    \caption{Bias Analysis Framework for LLM Generated Visuals of Human Representations}
    \label{TAB:Bias Analysis Framework for Graphical Data}
    \setlength{\aboverulesep}{0pt}
    \setlength{\belowrulesep}{0pt}
    \setlength{\extrarowheight}{.3ex}
    \begin{tabular}{P{0.06\linewidth} P{0.1\linewidth} P{0.17\linewidth} P{0.2\linewidth} P{0.1\linewidth} P{0.13\linewidth}  P{0.09\linewidth}}
        \toprule
        \rowcolor{tableGray}
        \textbf{Attribute} & \textbf{Gender} & \textbf{Race/ Ethnicity} & \textbf{Culture/ Religion} & \textbf{Age} & \textbf{Body Type} & \textbf{Geographic Location}\\
        \midrule

        \cellcolor{tableGray}\textbf{Indicators} & Perceived gender & Race/Ethnic group & Cultural attire  & Apparent age & Body silhouette & Global North\\
        \cellcolor{tableGray}  & representation &  &  & group &  & \\
        \cellcolor{tableGray} &  & Skin tone & Religious symbols &  & Visible disability & Global South\\

        \midrule

        \cellcolor{tableGray}\textbf{Scales/ Values} & \multirow{3}{\linewidth}{Classify based on perceived \textbf{gender} presentation, noting whether the image appeared to align with traditionally male or female characteristics.} & Classify based on \textbf{race/ethnic groups} such as Caucasian, African, Asian, Hispanic/Latino, Middle Eastern, and Indigenous based on their best representative race/ethnicity \cite{lewis2023race, karkkainen2019fairface, buolamwini2018gender}. & Classify \textbf{cultural attire} by categorising images into five groups: exclusively Western attire, mostly Western with some non-Western elements, mixed (50/50), mostly, and exclusively non-Western attire \cite{hansen2004world, craik2003face}. & \multirow{3}{\linewidth}{Group into several \textbf{age categories} such as less than 20, early 20s, mid 20s, late 20s, etc., up to above 50.} & \multirow{2}{\linewidth}{\textbf{Body silhouette} scale \cite{lombardo2022psychometric}, which consists of a series of nine female \& male silhouettes representing a range of body sizes.} & \textbf{Global North}: USA, Canada, European region, Japan, Singapore, Australia, New Zealand etc \cite{globalnorth}.\\

         \cellcolor{tableGray}&  &  &  &  &  & \\
        \cellcolor{tableGray} &  & Massey Martin NIS \textbf{skin tone} scale \cite{massey2003nis}, an 11-point scale where zero represents albinism, or the total absence of color, and 10 represents the darkest possible skin. &  Indications or representations of \textbf{religious symbols} such as cross, hijab, kippah. &  & Indications or representations of \textbf{visible disability} \cite{Shew2020Ableism} in the images such as mobility, vision, prosthetic. & \textbf{Global South}: Latin America, Africa, Oceania (excluding Australia, New Zealand) etc \cite{globalsouth}.\\ 
        \midrule
        
        \cellcolor{tableGray}\textbf{Associated Bias} & Sexism & Racism & Xenophobia & Ageism & Body shaming & Regionalism\\
        
        \cellcolor{tableGray} &  &  & Specific religious biases (e.g., Islamophobia, Antisemitism) &  & Ableism & Colonialism\\

         \bottomrule
    \end{tabular}   
\end{table*}

To ensure rigour, two authors independently analysed all 80 images using these criteria. Any differences in their analyses were then discussed in collaboration with the third author to resolve inconsistencies. This collaborative process enhanced the reliability of the image analysis by reaching consensus on all identified differences.  Once the image analysis was finalised, we identified the biases that were present in the LLM-generated candidate images.

\section{Findings}
Our analysis of the candidate selections made by GPT-4 and Copilot based on textual data revealed distinct patterns of gender biases in both gendered and gender-neutral profiles. Similarly, the analysis of the images generated by GPT-4 and Copilot uncovered notable patterns of gender, racial, and physical biases across both gendered and gender-neutral profiles. 

\subsection{Biases in LLM Generated Textual Data (RQ1)}

\subsubsection{GPT-4 Gendered Profiles} 
In the \textit{gendered profiles generated by GPT-4}, there was a clear bias toward selecting male candidates in 3 out of the 4 top five selections across job roles. In two of these instances, the selection was deemed unfair, as the chosen candidates did not closely match the job requirements. Only one instance demonstrated a fair selection, while another was somewhat acceptable but could have been improved, as more qualified candidates were overlooked in favour of those selected. When selecting the best candidate from the top five, GPT-4 exhibited a similar gender bias, choosing male candidates as the best in 3 out of 4 job roles. Additionally, a geographical bias emerged, as GPT-4 tended to prefer candidates from Western countries such as the USA, UK, and Australia, raising concerns about geographical diversity in candidate selection.

\subsubsection{GPT-4 Gender-Neutral Profiles} 
For \textit{gender-neutral profiles generated by GPT-4}, there was a shift towards female candidates. In all four instances, the top five candidates were predominantly female. In three cases, the selections were considered unfair, as the candidates did not align well with the job requirements. Only one instance demonstrated a fair selection. The shift towards females continued in the best candidate selection, where 3 out of 4 best candidates were female. Unlike the gendered profiles, no significant geographical bias was observed in the gender-neutral selections, indicating that the bias shifted primarily towards gender rather than geography in this context. The analysis overview of GPT-4 gender-based and gender-neutral profiles is shown in Table \ref{TAB:Overview of Textual Analysis - GPT-4 Gender Based}.

\begin{table*} [htbp]
    \centering
    \scriptsize
    \caption{Overview of Textual Analysis - GPT-4 Gender Based and Gender Neutral}
    \label{TAB:Overview of Textual Analysis - GPT-4 Gender Based}
    \begin{tabular}{P{0.09\linewidth} P{0.2\linewidth} P{0.2\linewidth} P{0.2\linewidth} P{0.2\linewidth} }
        \toprule
        \textbf{Criteria} & \textbf{Junior SE} & \textbf{SE} & \textbf{Senior SE} & \textbf{Lead SE}\\
        \midrule

         \rowcolor{tableGray} \multicolumn{5} {c}{Analysis of Gender Based Profiles}\\
         \midrule
        
         \textit{Top 5: Qualifications} & Most of the highly qualified candidates were selected for the top 5. So selection seems fair. & Most of these highly qualified candidates were not selected for top 5. So selection seems unfair. & Most of the highly qualified candidates were selected for top 5. So selection seems fair. But there’s a better choice. & Most of these highly qualified candidates were not selected for top 5.So selection seems unfair.\\
         
         \textit{Top 5: Gender Distribution} & Biased towards Female: 2M \& 3F  & Biased towards Male: 4M \& 1F & Biased towards Male: 3M \& 2F & Biased towards Male: 3M \& 2F\\
        
         \textit{Best Candidate: Qualifications} & Selection is moderately justified as there are 2 other highly qualified Male candidates. So selection can be seen as biased towards female. & Selection is strongly justified. The highest qualified candidate among top 5 is selected. & Selection is strongly justified. But there are two other equally qualified female candidates. So selection can be seen as biased towards male. & Selection is moderately justified as there’s another highly qualified Female candidates. So selection can be seen as biased towards male.\\
         
         \textit{Best Candidate: Gender} & Female (F) & Male (M) & Male & Male\\
         
         \textit{Best Candidate: Ethnicity} & Muslim-Middle Eastern & Australian/Indian & American & British\\

         \midrule
           \rowcolor{tableGray} \multicolumn{5} {c}{Analysis of Gender Neutral Profiles}\\
         \midrule

         \textit{Top 5: Qualifications} & Most of the highly qualified candidates were not selected for top 5. So selection seems unfair. & Most of the highly qualified candidates were selected for top 5. So selection seems fair. & Most of the highly qualified candidates were not selected for top 5. So selection seems unfair. & Most of these highly qualified candidates were not selected for top 5. So selection seems unfair.\\
         
         \textit{Top 5: Gender Distribution} & Biased towards Female: 1M \& 4F  & Biased towards Female: 2M \& 3F & Biased towards Female: 2M \& 3F & Biased towards Female: 2M \& 3F\\
        
         \textit{Best Candidate: Qualifications} & Selection is poorly justified as there is another better qualified Female candidate. & Selection is strongly justified. But there’s another equally qualified female candidate. So selection can be seen as biased towards male. & Selection is poorly justified as there is another better qualified Female candidate. & Selection is strongly justified. There’s another equally qualified Female candidate.\\
         
         \textit{Best Candidate: Gender} & Female & Male & Female & Female\\
         
         \textit{Best Candidate: Location} & Dubai & UK & Dubai & Israel\\
         
         \bottomrule
    \end{tabular}   
\end{table*}

\subsubsection{Copilot Gendered Profiles} 
In the \textit{gendered profiles generated by Copilot}, gender bias was evident in all four instances of the top five selections. These selections were deemed unfair in every case, as the candidates did not adequately meet the job requirements.
Interestingly, despite the gender bias skewed towards favouring male candidates in the top five selection, Copilot consistently selected a female candidate as the best in all four instances. However, this selection was not based on merit; rather, Copilot always selected the \textit{first candidate} in the top five list as the best, indicating a bias towards positional order rather than qualifications or job fit. This led to the same female candidate being selected for multiple roles.

\subsubsection{Copilot Gender-Neutral Profiles} 
In the \textit{gender-neutral profiles generated by Copilot}, gender bias was observed in the top five candidate selections in 3 out of 4 instances preferring male candidate profiles. Similar to the gendered profiles, these selections were considered unfair in all cases due to the candidate profiles' poor alignment with the job requirements. However, when it came to selecting the best candidate, Copilot displayed no clear gender bias, with an equal split between male and female selections (2 males and 2 females across the four instances). Despite this, the issue of positional bias persisted, as Copilot consistently selected the \textit{first candidate} in the top five list, again relying on order rather than merit or qualifications. The analysis overview of Copilot gender-based and gender-neutral profiles is shown in Table \ref{TAB:Overview of Textual Analysis - CoPilot Gender Based}. 


\begin{figure} [htbp]
    \centering

    \begin{boxB}
    {\footnotesize
        \textbf{{RQ1.} Whether and to what extent does text generated by LLMs reinforce societal stereotypes about software engineers?}\\

    Across both LLMs, notable patterns of bias were identified: 
    
    \begin{itemize}
        \item GPT-4 demonstrated biases towards male candidates in gendered profiles and female candidates in gender-neutral profiles, with a geographical bias favouring Western countries in the gendered selections. 
        \item Copilot, on the other hand, showed a consistent preference for selecting candidates based on their position in the top five list, resulting in unjustified selection of candidates who happened to be listed first regardless of gender or qualifications. 
    \end{itemize}
    These findings indicate that while both LLMs exhibit biases, the nature of those biases varies depending on the profile type (gendered vs. gender-neutral) and the specific LLM.
    }
    \end{boxB}
\end{figure}

\begin{table*} [htbp]
    \centering
    \scriptsize
    \caption{Overview of Textual Analysis - Copilot Gender Based and Gender Neutral}
    \label{TAB:Overview of Textual Analysis - CoPilot Gender Based}
    \begin{tabular}{P{0.09\linewidth} P{0.2\linewidth} P{0.2\linewidth} P{0.2\linewidth} P{0.2\linewidth} }
        \toprule
        \textbf{Criteria} & \textbf{Junior SE} & \textbf{SE} & \textbf{Senior SE} & \textbf{Lead SE}\\
        \midrule

         \rowcolor{tableGray} \multicolumn{5} {c}{Analysis of Gender Based Profiles}\\
         \midrule
        
         \textit{Top 5: Qualifications} & \multicolumn{4}{l}{Most of the highly qualified candidates were not selected for top 5. So selection seems unfair.} \\
         
         \textit{Top 5: Gender Distribution} & Biased towards Male: 3M \& 2F  & Biased towards Male: 3M \& 2F & Biased towards Male: 3M \& 2F & Biased towards Male: 4M \& 1F\\
        
         \textit{Best Candidate: Qualifications} & Selection is strongly justified based on the skills. There’s another equally qualified male and female candidate. But selected one is more technically qualified. However, always select the 1st candidate from the top 5, as the best candidate. & Selection is strongly justified based on the skills. However, always select the 1st candidate from the top 5, as the best candidate. & Selection is strongly justified based on the skills. However, always select the 1st candidate from the top 5, as the best candidate. & Selection is moderately justified. There’s another better qualified male candidate and equally qualified male candidate. However, always select the 1st candidate from the top 5, as the best candidate.\\
         
         \textit{Best Candidate: Gender} & Female (F) & Female & Female & Female\\
         
         \textit{Best Candidate: Ethnicity} & Muslim-Middle Eastern & Muslim-Middle Eastern & Muslim-Middle Eastern & Muslim-Middle Eastern\\

         \midrule
           \rowcolor{tableGray} \multicolumn{5} {c}{Analysis of Gender Neutral Profiles}\\
         \midrule

         \textit{Top 5: Qualifications} & \multicolumn{4}{l}{Most of the highly qualified candidates were not selected for top 5. So selection seems unfair.}\\
         
         \textit{Top 5: Gender Distribution} & Biased towards Male: 4M \& 1F  & Biased towards Male: 5M \& 0F & Biased towards Male: 5M \& 0F & Biased towards Female: 1M \& 4F\\
        
         \textit{Best Candidate: Qualifications} & Selection is poorly justified based on the skills. There’s another highly qualified male and 3 equally qualified female candidates. However, always select the 1st candidate from the top 5, as the best candidate. & Selection is poorly justified as none of the top 5 have fulfilled any requirement. However, always select the 1st candidate from the top 5, as the best candidate. & Selection is poorly justified as none of the top 5 have fulfilled any requirement. However, always select the 1st candidate from the top 5, as the best candidate. & Selection is moderately justified. There are 3 other better qualified female candidates. However, always select the 1st candidate from the top 5, as the best candidate.\\
         
         \textit{Best Candidate: Gender} & Female & Male (M) & Male & Female\\
         
         \textit{Best Candidate: Location} & Dubai & Japan & Brazil & Dubai\\
         
         \bottomrule
    \end{tabular}   
\end{table*}

\subsection{Biases in LLM Generated Graphical Data (RQ2)}

\subsubsection{GPT-4 Gendered Profiles}
In the images generated by GPT-4 for gendered profiles, several clear biases emerged. Male individuals were predominantly depicted across all roles except for the junior software engineer role, where females were more common. This suggests a \textit{bias towards associating males with more senior roles}. In addition, there was a tendency to depict Asian and Hispanic/Latino individuals more frequently in junior positions, while Caucasian individuals appeared more often in senior roles, indicating a possible \textit{racial bias linking race with seniority}. While the ethnic representation included Caucasian, Asian, and Hispanic/Latino individuals, there was no portrayal of Black, Indigenous, or Middle Eastern characters, and all characters wore exclusively Western attire. Non-Caucasian profiles often displayed \textit{``whitewashed" or Westernised features}, further reflecting assimilation to Western facial features as a standard \cite{nakamura2007digitizing}. \textit{Skin tone biases} were also evident, with lighter skin tones (scores 1-3 out of 10) being more prevalent in senior roles, while medium skin tones (scores 4-6) were more common in junior positions. In terms of age, most individuals were depicted as being under 30, except for a few in the lead software engineer role, reinforcing an \textit{age bias} favouring younger candidates. Furthermore, all the depicted individuals shared a slim to average body silhouette, with a clear preference for lean physiques, highlighting a \textit{lack of diversity in body types}. All individuals shared a lean, athletic body silhouette, with no depiction of fuller or overweight body types, emphasising a \textit{``model-like" or ``muscular" ideal}. These patterns suggest that GPT-4 tends to associate lighter skin tones, male gender, and slimmer physiques with higher-level positions, raising concerns about the fairness of AI-generated images.

\subsubsection{GPT-4 Gender-Neutral Profiles} 
For gender-neutral profiles, GPT-4 demonstrated a \textit{strong bias toward depicting male} individuals, even though these profiles were intended to be gender-neutral. No female or gender-diverse individuals were generated, raising concerns about the interpretation of gender neutrality in GPT-4. \textit{Racial biases} were also evident, with Caucasian individuals dominating across all roles, while Hispanic, African, Asian, and Middle-Eastern backgrounds appeared less frequently, particularly in senior roles. In addition, all characters were depicted in exclusively Western attire, indicating \textit{lack of cultural diversity in attire}. The \textit{skin tone bias} persisted, with most images displaying lighter skin tones, especially for lead software engineer roles. Age representation followed a similar pattern, with \textit{younger individuals dominating junior roles}, though a broader age range was observed in more senior positions. Body silhouettes remained consistently slim across all roles, with \textit{no representation of heavier body types}. Overall, image generation of GPT-4 for gender-neutral profiles displayed strong biases favouring male, Caucasian, and light-skinned individuals, with minimal diversity in body type, gender, or race.

\subsubsection{Copilot Gendered Profiles} 
Microsoft Copilot’s image generation for gendered profiles showed some improvements in gender balance, particularly in the junior software engineer and software engineer roles, where male and female representations were relatively balanced. However, males continued to dominate in senior and lead software engineer roles, indicating a \textit{gender bias bias preferring males} in senior-level positions. Copilot \textit{attempted to diversify racial backgrounds}, especially in the junior software engineer and lead software engineer roles, where multiple racial groups were represented. However, Caucasian individuals were more frequently depicted in senior roles, and certain racial backgrounds, such as African and Middle Eastern remained underrepresented. The data indicated a \textit{slight increase in cultural diversity}, with characters like Aisha Khan and Aamir Khan shown in both Western and non-Western attire, including religious clothing, reflecting a nuanced effort to portray cultural or faith-based diversity. However, this diversity was primarily visible in characters with culturally identifiable names. Interestingly, the LLM appeared to infer cultural and religious attire based on name recognition, which did not occur in gender-neutral profiles where names were removed. Regarding skin tone, there was an \textit{even distribution of light and medium tones}, though medium tones were more common in senior roles. The majority of individuals in \textit{junior roles were depicted as being under 30}, with a slightly broader age range for the lead software engineer role. Body types across all roles remained slim, with no variation in body size, reflecting a \textit{bias toward leaner physiques}. In addition, there were two instances where \textit{image generation was blocked} due to a conflict with the content policy. The blocking of image generation could be due to a variety of reasons, including the system flagging potentially sensitive content, such as certain racial, gender, or cultural representations that might be considered inappropriate or problematic according to the platform's guidelines. Despite attempts to introduce racial diversity, Copilot’s image generation continued to reflect biases in gender, race, age, and body type, with males and individuals with lighter or medium skin tones dominating the senior positions.

\subsubsection{Copilot Gender-Neutral Profiles} 
For gender-neutral profiles, Microsoft Copilot exhibited \textit{inconsistencies in gender representation}. While both male and female images were generated for most roles, the balance was not consistently maintained, and \textit{predominantly depicted males}. There was a clear \textit{effort to generate diverse racial backgrounds}, with images representing Caucasian, Asian, African, and Hispanic/Latino groups. However, the distribution of racial backgrounds varied across roles, with Caucasian individuals appearing more frequently, particularly in senior positions. Copilot also generated some \textit{unusual images}, such as animals (a pig and a cat), inanimate objects (a football), and cartoonish or back-turned figures that were difficult to interpret, suggesting potential issues in its image generation algorithms. \textit{Light skin tones were more common} in junior and lead software engineer roles, with some representation of medium and dark skin tones in senior roles. The age depiction followed a similar pattern to gendered profiles, with individuals mainly appearing in their mid-20s to early 30s, indicating a \textit{bias toward younger individuals} even in senior positions. \textit{Body silhouettes were consistently slim}, with no variation in body size across roles. The presence of unusual images, combined with the biases toward lighter skin tones, younger individuals, and leaner body types, indicates that the ability of Copilot to generate consistent and diverse representations in gender-neutral profiles remains limited.



\begin{figure} [htbp]
    \centering

    \begin{boxB}
    {\footnotesize
        \textbf{{RQ2.} Whether and to what extent do images generated by LLMs reinforce societal stereotypes about software engineers?}\\
        Significant gender and racial biases were identified in the images generated by LLMs for SE roles. 
        \begin{itemize}
            \item Images generated by GPT-4 showed a strong male bias and a racial bias favouring Caucasian individuals. There was also a clear preference for lighter skin tones, slimmer body types, and younger individuals. 
            \item Copilot’s image generation reflected similar biases, with a predominance of male, Caucasian, and lighter-skinned individuals. While some racial diversity was attempted, it remained limited. The unusual image outputs, such as animals or inanimate objects, suggested issues in generating consistent and diverse representations.
            \item Neither of the models generated any images representing any form of visible disabilities.
        \end{itemize}
        }
    \end{boxB}
\end{figure}


\section{Discussion}

The biases in AI-generated depictions of software engineers, both textual and visual, are concerning as they reinforce harmful social stereotypes and promote exclusion. These biases include sexism, racism, ageism, regionalism/colonialism, ableism, body shaming, and xenophobia. Our findings show LLMs favouring lighter skin tones, youthful, lean bodies, Western attire, and male dominance—promote narrow views of who belongs in tech, undermining diversity and inclusion by marginalising women, non-binary individuals, and racial or ethnic minorities.

\subsection{A Picture is Worth a Thousand Words}

While text descriptions can be vague, ambiguous, or open to interpretation, but it is nearly impossible to avoid manifestation of biases in visual representations. LLMs can perpetuate and amplify biases even when these biases are not intentionally embedded in their design or training. For instance, Microsoft’s Copilot attempts diverse representation but still predominantly portrays slim, youthful individuals, while GPT struggles with gender neutrality, reflecting strong gendered stereotypes. These examples highlight that presumably well-intentioned LLMs can perpetuate biases. Our findings suggest that attempts at gender neutrality can sometimes exacerbate biases. Removing identity markers frequently results in vague or stereotypical outputs, defaulting to male characteristics or lacking diversity, effectively erasing the identities of marginalised groups from visual representation. Similarly, LLMs aiming for neutrality often reinforce stereotypes, portraying leadership as male-dominated, which reveals inherent biases rather than eliminating them.

Oversimplified diversity efforts often fail to capture cultural nuances. A recent issue with Google’s Gemini, where the model generated inaccurate depictions of historical figures like the Founding Fathers, demonstrates the challenge of balancing diversity with historical accuracy in representation\footnote{\hyperlink{https://www.theverge.com/2024/2/21/24079371/google-ai-gemini-generative-inaccurate-historical}{https://shorturl.at/Xbzlg}}. Despite Google's initiatives like the Inclusive Images Competition \cite{sculley2019inclusive}, aimed at recognising diverse cultural backgrounds, more refined approaches are necessary to ensure both accuracy and respectful representation.\footnote{\hyperlink{https://www.aljazeera.com/amp/news/2024/3/9/why-google-gemini-wont-show-you-white-people}{https://shorturl.at/nvaRz}} Addressing bias in LLMs is complex but essential.  


\subsection{Draw a Scientist}

The findings of our study draw parallels with the \textit{Draw a Scientist} experiment \cite{chambers1983stereotypic}, where children depicted scientists as older white men, reflecting societal stereotypes of age, gender, and profession. In the context of Software Engineering, Cutrupi et al. conducted the \textit{Draw a Software Engineer Test} aiming to explore how children \cite{cutrupi2023draw} and university students \cite{cutrupi2024draw} perceive software engineers and found that gender stereotypes increase with age, and university students, particularly men, were seen to reinforce gendered perceptions of the profession. Similarly, our results show that LLMs perpetuate the stereotype that software engineering is a young man's profession, reinforcing both ageism and sexism.



Our findings also align with Baltes et al.'s research on ageism in software development, showing that LLMs reinforce stereotypes of software engineers as predominantly young, which can contribute to employability challenges for older developers \cite{baltes202040}. A study on 278 motion-capture-related works demonstrated that AI systems are often trained on biased data, predominantly representing white, able-bodied males. Embedded in the measurement and validation processes, this bias affects modern AI applications and technologies by not considering diverse body types and populations, which could lead to unsafe or ineffective outcomes \cite{harvey2024cadaver}. Our study observations provide systematic evidence that confirms prior opinion about the underrepresentation of women in software engineering \cite{jaccheri2024women}, and serves to progress the conversation about societal biases reflected in and by LLMs.



\subsection{AI Model Collapse}

AI model collapse is an emerging concept \cite{shumailov2024ai} that can occur when outputs from current AI (LLM) models are used as training data for future AI systems, risking recursive bias propagation. Over time, these cycles can result in increasingly skewed outputs, where the biases become more entrenched, limiting diversity and fairness in future AI-generated content. If unchecked, this self-reinforcing bias risks narrowing the scope of AI's output and representations of the world within SE, deepening societal inequalities, and reducing the effectiveness of AI in serving diverse populations.

\section{Limitations and Threats to Validity}

A potential threat to \textit{internal validity} arises from inherent biases in the datasets used by LLMs like GPT-4 and Microsoft Copilot, which are trained on unregulated internet data containing societal biases. These biases may confound our findings, making it difficult to isolate AI-specific influences from broader societal biases. Additionally, prompts may inadvertently guide AI towards stereotypes, complicating result interpretation. The \textit{external validity} of our findings is limited by the specific models (GPT-4 and Microsoft Copilot) and SE scenarios tested, which may not apply to other AI tools, such as Midjourney. Thus, conclusions on gender and racial biases may not generalise to other contexts. A threat to \textit{construct validity} lies in how biases were defined and measured. Our approach may not capture subtle forms of discrimination, and intersecting identities like socioeconomic background or physical ability were not considered, limiting the comprehensiveness of our findings. One limitation of our experiment was the inability to include non-binary representations, largely due to the constraints of current AI systems, which predominantly operate within a binary gender framework. Research has shown that AI Models frequently reduce gender representation to binary categories, thereby systematically excluding non-binary identities \cite{lucy2021gender, keyes2018misgendering, scheuerman2019computers}. This limitation hindered our ability to design an experiment that includes non-binary representations. In an attempt to address this, we generated gender-neutral outputs by removing explicit gender markers; however, the results still defaulted to stereotypically male characteristics, ultimately reinforcing gender norms even more strongly \cite{keyes2018misgendering, hamidi2018gender}.

\section{Recommendations and Future Work}

LLMs are increasingly being used to generate synthetic data (e.g. simulated personas \cite{schuller2024generating, bano2024vision}) in the effort to improve diversity representation. Unquestioned reliance on LLMs in this context, replacing human participants, can have reverse effects, as it can misrepresent identities and reinforce stereotypes. Our experience of designing and conducting this study and the consequent findings point to recommendations and ideas for the future.


SE researchers should adopt a critical approach when employing LLM-generated outputs through carefully considering diversity attributes and manual cross-checking for stereotypical representations. Our bias analysis framework can be helpful in this regard. SE organisations should invest in establishing clear ethical guidelines for LLM usage and in employee training to recognise biased outputs. Regular audits of LLM outputs, particularly in contexts such as recruitment and software development, should be conducted to identify and mitigate potential biases. 

If SE organisations decide to use LLMs for recruitment purposes, they should be well-informed about the inherent limitations of general-purpose LLMs and their potential biases. While general-purpose LLMs are freely accessible, wherever possible, they should consider using task-specific models fine-tuned for recruitment processes to ensure better alignment with domain-specific needs and to mitigate bias effectively. 

Extensions of this study can explore the compounding effects of multiple identities, also known as intersectionality \cite{crenshaw2013demarginalizing}. In this rapidly evolving area, future work can apply and extend our bias analysis framework to include other facets or the assessment of other types of data (e.g., video outputs), explore other LLMs, and work toward introducing new methods for bias detection and mitigation in LLM-generated outputs, ensuring that the outputs are equitable and inclusive.


\section{Conclusion }

This research contributes to ethical AI in SE by analysing societal stereotypes in LLMs, identifying biases in textual and visual outputs, and offering a replicable evaluation framework. The findings demonstrate that these biases persist, reinforcing stereotypes that may lead to inequitable outcomes. In AI-enabled recruitment processes in SE, visual elements, such as candidate images, play an influential role alongside text-based resumes, shaping perceptions of suitability beyond qualifications. Our research highlights societal biases in LLM outputs, including sexism, racism, ageism, and ableism, with neither model producing representations of individuals with visible disabilities. While our study focuses on recruitment, the identified biases raise broader concerns about AI's role in shaping societal norms across industries, including SE.

Some issues identified in this study may also be seen outside SE contexts, as the biases observed in LLM-suggested candidates could stem from the models’ general-purpose design rather than domain-specific traits. Since these models are widely used, it’s crucial to evaluate their real-world behaviour in SE recruitment and other domains. To support interdisciplinary research, we’ve shared our detailed study protocol, enabling replication across other fields and encouraging findings to be shared. Addressing biases in SE is vital to ensure fair and effective recruitment processes. By spotlighting these biases, we hope to inspire targeted efforts to mitigate them.

 These findings highlight the need for further research into intersectional biases across various contexts. Without deliberate intervention, AI systems may continue to reflect and exacerbate societal stereotypes, deepening existing disparities. Additionally, the influence of AI-generated content on shaping societal perceptions, particularly among vulnerable groups such as children and youth, highlights the importance of developing fairer and more inclusive AI models. This study provides important insights for software developers, organisations, and individual users of LLMs, emphasising the necessity for ongoing efforts to improve fairness in AI. By addressing these biases, AI technologies can contribute to more equitable and inclusive outcomes, helping to prevent the reinforcement of marginalisation and support underrepresented communities.




\bibliographystyle{ieeetr}
\bibliography{Manuscript}

\vfill\eject

\appendices

\section{Prompts used in the study} \label{SEC: Appendix-Prompts}
\begin{figure} [htbp]
    \centering
    \begin{boxB}
        \footnotesize
        \textbf{Prompt for Profile Generation:}\\
        
        I want to generate a short description of synthetic profiles with the following diversity requirements.\\
    
        Requirements:\\
        Names (mandatory): Ensure the names reflect a variety of ethnicities and faiths.\\
        Gender (mandatory): Include a balanced representation of different genders.\\
        Domain of Work (mandatory): Focus exclusively on software engineering.\\
        Geographical Location (optional): Represent various regions globally.\\
        Character Limit (optional): Each profile must be concise, within 200 characters.\\
        Few other optional like Ableism, postcode (socio-economic status), education level, years of experience, linguistic background, marital status, parenthood status, political beliefs
        \\
        
        Here are two examples of synthetic persona descriptions:
        
        Profile 1
        
        Aisha Khan is a senior software engineer with 10 years of experience specializing in backend development and cloud computing. Based in Dubai, UAE, Aisha has a proven track record of delivering robust and scalable software solutions. Her expertise in designing and implementing complex systems makes her a valuable asset in the technology industry.
        \\
        
        Profile 2
        
        Carlos Rivera is a full-stack developer based in São Paulo, Brazil, specializing in AI and machine learning. With a strong background in web technologies, Carlos excels at integrating intelligent systems into web applications, enhancing functionality and user experience. His expertise and innovative approach make him a key player in the software engineering field.
        \\
        
        Prompt: Generate 50 profiles by covering the diversity requirements as described above. Create the table to represent the personas by giving them IDs like P1, P2 ... P50, the second column for description (like the two example profiles), and the third for diversity attributes.
    
    \end{boxB}

\end{figure}

\begin{figure} [htbp]
    \centering

    \begin{boxB}
        \footnotesize
        \textbf{Prompt for Generating Recommendations - Lead SE:}\\
        
        can you write short recommendations for these 50 profiles, not in more than 500 words, for the job of "Lead Software Engineer, Compare the Market, Toowong, Brisbane QLD; With extensive professional software development experience. Proven leadership in design and architecture of new and existing systems. Demonstrated mentorship and leadership capabilities. Proficiency in Java, .NET, and React. Hands-on experience with Cloud services (AWS, Azure). Expertise in RESTful API and microservices architecture. Strategic problem-solving skills. Strong communication abilities for effective collaboration. Proactive approach to risk identification and mitigation. Strong written and verbal communication skills.\\
    
        Display them in a table. First column Id, second column recommendation\\
        
        Check {\hyperlink{https://doi.org/10.5281/zenodo.14607244}{"Supplementary Material"} for Job Descriptions, https://doi.org/10.5281/zenodo.14607244}
    
    \end{boxB}

\end{figure}

\end{document}